\documentclass[12pt]{article}
\usepackage{array}
\usepackage{graphicx}
\usepackage{amssymb}
\usepackage{amsmath}
\input{epsf}
\usepackage{cite}
\def\@fmsl@sh#1#2#3{\m@th\ooalign{$\hfil#1\mkern#2/\hfil$\crcr$#1#3$}}
 \def\eq#1\en{\begin{equation}#1\end{equation}}
\def\s[#1,#2]{[#1\stackrel{\star}{,}#2]}
\def\sx[#1,#2]{[#1\stackrel{\star_{x}}{,}#2]}

\newcommand{\nc}{\newcommand}
\nc{\beq}{\begin{equation}}
\nc{\eeq}{\end{equation}}
\nc{\beqa}{\begin{eqnarray}}
\nc{\eeqa}{\end{eqnarray}}

\def\bc{\begin{center}}
\def\ec{\end{center}}

\def\to{\rightarrow}

\def\gsim{\mathrel{\mathpalette\atversim>}}

\def\bc{\begin{center}}
\def\ec{\end{center}}

\def\gsim{\mathrel{\rlap{\lower4pt\hbox{\hskip1pt$\sim$}}

    \raise1pt\hbox{$>$}}}       

\def\gsim{\mathrel{\rlap{\lower4pt\hbox{\hskip1pt$\sim$}}
    \raise1pt\hbox{$>$}}}       

\begin{document}
\makeatletter
\def\fmslash{\@ifnextchar[{\fmsl@sh}{\fmsl@sh[0mu]}}
\def\fmsl@sh[#1]#2{%
  \mathchoice
    {\@fmsl@sh\displaystyle{#1}{#2}}%
    {\@fmsl@sh\textstyle{#1}{#2}}%
    {\@fmsl@sh\scriptstyle{#1}{#2}}%
    {\@fmsl@sh\scriptscriptstyle{#1}{#2}}}
\def\@fmsl@sh#1#2#3{\m@th\ooalign{$\hfil#1\mkern#2/\hfil$\crcr$#1#3$}}
\makeatother

\thispagestyle{empty}
\begin{titlepage}
\boldmath
\begin{center}
  \Large {\bf Lower Dimensional Quantum Black Holes}
    \end{center}
\unboldmath
\vspace{0.2cm}
\begin{center}
{
{\large Xavier Calmet}\footnote{x.calmet@sussex.ac.uk}$^a$
{\large  and}
{\large Greg Landsberg}\footnote{landsberg@hep.brown.edu}$^{b}$
}
 \end{center}
\begin{center}
{\sl $^a$Physics and Astronomy, 
University of Sussex,   Falmer, Brighton, BN1 9QH, UK 
}\\
{\sl  $^b$Department of Physics, Brown University, Providence, RI 02912, USA}
\end{center}
\vspace{\fill}
\begin{abstract}
\noindent
There are theoretical reasons to think  that space-time is lower dimensional at very short distances. If this is the case, quantum black holes produced at the LHC or in cosmic rays scattering live in lower dimensions. We discuss production cross section and signatures for the corresponding quantum black holes at the LHC within several different models of low-dimensional space-time.
\end{abstract}  

\end{titlepage}



\newpage

Models with low-scale quantum gravity (see, e.g. \cite{Calmet:2010nt} for a recent review) have received a considerable amount of attention over the last decade. One of the most striking signatures of these models is the possibility of producing mini black holes at colliders~\cite{Dimopoulos:2001hw,Giddings:2001bu}.

The formation of a classical black hole in a collision of two particles with zero or non-zero impact parameter is remarkably well understood, thanks to the beautiful work by Penrose, unfortunately unpublished, which was followed by D'Eath and Payne~\cite{D'Eath:1992hb} and more recently by Eardley and Giddings~\cite{Eardley:2002re}. The construction of Penrose is valid in the limit where the collision energy, and thus the mass of the black hole,  is much larger than the Planck mass. This seminal work establishes that the production cross section for black holes in the collision of two particles at very high energy is given by $\sigma \approx \pi r_s^2$ where $r_s$ is the Schwarzschild radius, which depends on the mass of the black hole (and hence of the center-of-mass energy) and on the dimensionality of space.

An extension of this construction in the semi-classical regime was proposed by Hsu~\cite{Hsu:2002bd} using a path-integral approach. This construction holds if there is a hierarchy between the Planck mass and the center-of-mass energy. One typically estimates that the semi-classical black hole mass is some 5 to 20 times larger than the Planck mass. For this reason, it is now well understood that the likelihood of producing semi-classical black holes at the LHC, which are thermal objects decaying into a multiparticle final state, is quite small.

Whereas it is hard to produce a semi-classical black hole at the LHC, quantum black holes (QBHs), which we define as non-thermal mini black holes with masses comparable to the Planck mass \cite{Calmet:2008dg}, could be produced copiously. Unfortunately, very little is known about QBHs since their description would require a complete understanding of quantum gravity. The physics of QBHs can be, to a certain extent, extrapolated from the understanding we have of the semi-classical black holes. If  one considers the limit where the mass of a semi-classical black hole decreases, i.e., its temperature then increases, the black hole becomes less and less thermal, and the number of the final-state particles decreases as a result.

A truly quantum black hole is expected to decay into only a handful of particles, which would resemble strong gravitational rescattering.  QBHs can be thought of as small gravitational bound states, which are short-lived. At a proton-proton collider, QBHs will be produced in the collisions of quarks and gluons. Therefore most of the time they will carry color and electric charges and can be classified according to the representations of  the groups $SU(3)_{\mbox{color}}$ and $U(1)_{\mbox{QED}}$, since gauge symmetries are expected to be conserved by quantum gravity. Note that this does not conflict with confinement since the typical length scale of QCD, $\Lambda_{QCD}^{-1}$, is much larger than the Planck length $\sim M_P^{-1}$, where $M_P$ is the Planck mass. This allows one to make rather precise predictions for the LHC. However, one crucial piece of information is difficult to obtain: it is that of the cross section of the QBH formation. In the past, this cross section has been extrapolated from the semi-classical approximation as well. This could be too na\"ive. 

We do not have much knowledge of the structure of space-time are very short distances, i.e. at the energies relevant to the formation of QBHs. However, it has recently been argued by Ambjorn, Jurkiewicz, and Loll~\cite{Ambjorn:2004qm} using a causal dynamical triangulations approach to quantum gravity,  which is a modification of quantum Regge calculus where space-time is discretized, that space-time is 1+1-dimensional at very short distances. There is a clear connection of this idea with the models proposed by Horava~\cite{Horava:2009uw} and more recently by one of us and collaborators~\cite{Anchordoqui:2010er}.

If space-time is truly 1+1-dimensional at very short distances, then QBHs are expected to live in 1+1 dimensions. Gravity in the two-dimensional space-time can be described by the following action (see, e.g. \cite{Callan:1992rs,Lemos:1994fn}):
\begin{eqnarray} \label{action2d}
S= \frac{1}{2 \pi} \int d^2 x \sqrt{-g}  e^{-2 \phi} ( R - 4 w  (\nabla \phi)^2 + 4 \Lambda^2),
\end{eqnarray}
where $w$ is a dimensionless parameter, $g$ is the metric, $\phi$ is the dilaton field, and $\Lambda$ is the cosmological constant. Note that $w=-1/2$ corresponds to planar General Relativity  and $w =- 1$ to the first-order string theory action. The limit $w \to \infty$ is the closest analog to General Relativity in 1+1 dimensions \cite{Lemos:1994fn}. Besides the cosmological constant, there are no dimensional parameters in this action. Classical black hole solutions are known in this theory. However, here we are concerned with non-thermal quantum black holes.  If the space-time is only two-dimensional at very short distances of the order of the inverse Planck mass, there are important implications for the production cross section of QBHs, namely the cross section has to be dimensionless since it corresponds to 1+1-dimensional physics. This two-dimensional cross section is either a constant ($\sigma_2 = \mbox{constant}$) or a function of the ratio of the center-of-mass energy to the cosmological constant, which is the only other dimensionless parameter one can build from the variables in the action ($\sigma_2 = f(\sqrt{\hat s}/\Lambda))$.   We need to convert the two-dimensional cross section to the one which is measured at macroscopic distances, in 3+1 dimensions. The important physical quantity is the flux of particles contained in the cylinder of diameter $\sim b$, where $b$ is the impact parameter of the scattering between the two particles. To achieve the correct dimensionality, the four-dimensional cross section is then $\sigma_4 \sim \sigma_2 b^2\theta(\sqrt{\hat s}-b^{-1})$.

Let us now discuss different models of space-time at very short distances. We consider the gravitational scattering of two particles in a collider. We shall first assume that gravity becomes strong at 1 TeV. Strong gravitational physics is felt by the colliding particles if $b^{-1} \sim M_P$ where $M_P$ is the effective Planck mass of the model. This could be realized in models with large extra dimensions or a large hidden sector, see, e.g.  \cite{Calmet:2010nt} for a recent review. Two scenarios are possible. The first one is that quantum black holes never form because the parameters of the action (\ref{action2d}) are such that gravity at very short distances is weak and does not have a minimal length. This theory is likely to be pathological as it would imply that there are real singularities in astrophysical black holes or during the Big Bang. If this was the case, semi-classical black holes would still be formed at high energy, but when particles collide at center-of-mass energies similar to the Planck mass, they would just fly by another without interacting gravitationally.  We could call this model asymptotically-free gravity in 1+1 dimensions.

A second, more exciting possibility is that the parameters of the action  (\ref{action2d}) are such that gravity is strong as well in the 1+1-dimensional regime. In that case we expect quantum black holes to form in collisions at center-of-mass energies of the order of $\sqrt{\hat s}\sim b^{-1}\sim M_P$.
We expect that the correct 4D cross section for quantum black holes should be a limit of the semi-classical black hole cross section:
\begin{eqnarray}
\sigma= a_n \pi r_n^2,
\end{eqnarray}
where $n$ is the number of extra dimensions in space at the macroscopic distances, $a_n$ is a calculable numerical factor of order one and, $r_n$ is the $4+n$ Schwarzschild radius given by:
\begin{eqnarray}
r_s(\hat s,n,\bar M_P)=\left [\frac{ \Gamma\left ((3+n)/2 \right )}{(2+n)} 2^n \sqrt{\pi}^{n-3} \ \frac{\sqrt{\hat s}}{\bar M_P}\ \right ]^\frac{1}{1+n}  \frac{1}{\bar M_P},
\end{eqnarray}
where $\bar M_P$ is the reduced Planck mass.  
 We  propose the following cross section for QBHs from the four-dimensional point of view:
\begin{eqnarray}
\sigma_{QBH}= \frac{1}{ 16 \pi \bar M_P^2} \theta({\sqrt{\hat s} -\bar M_P}).\label{eq:cs}
\end{eqnarray}
This cross section can be obtained from the one for classical black holes by taking the limit $\sqrt{\hat s} \to \bar M_P$. Note that the cross section is universal and does not grow with the quantum black hole mass. This ansatz is particularly interesting since it might provide a solution to the unitarity problem of General Relativity, since this cross section does not grow with energy. The step function implies that the two colliding partons must be within a distance of $\bar M_P^{-1}$ from one the other to form a quantum black hole.

We are assuming that the production cross section for QBHs is independent of the number of dimensions of space-time at intermediate energy scales $\sim M_P$. Indeed, there are different frameworks in which lower-dimensional quantum black holes  could be produced. The most obvious one is that of a four-dimensional theory with a large hidden sector \cite{Calmet:2010nt}. In the case of large extra dimensions at the macroscopic scale, this scenario is realizable if large extra dimensions open up at short distances, before collapsing to two dimensions in the full quantum-gravity regime.  Since space and time are dynamical objects, this is easily imaginable. This is why we posit a universal cross section, which is independent of the number of dimensions.

Quantum black holes \cite{Calmet:2008dg}, in contrast to semi-classical ones, are very short-lived non-thermal objects and will thus decay immediately into a few particles, typically just two. As suggested by Meade and Randall~\cite{Meade:2007sz}, the corresponding physics would resemble strong gravitational scattering. As we have pointed out, this would take place in 1+1 space-time dimensions. Clearly this will affect the way one searches for QBHs at the LHC.  The angular distribution of the two particles produced by the decaying  black hole is model dependent. If there is no space-time foam, we can imagine that the particles colliding with energies close to the Planck mass will generate gravitational fields, which will resemble that of the Aichelburg-Sexl metric~\cite{Aichelburg:1970dh}, i.e., they will look like shock waves. In that case, the angular distribution is very peculiar. A QBH will decay into two particles back-to-back, which will in most cases form two energetic jets.  These jets will essentially be in the direction of the colliding beams and as such very tough to see. We still expect a suppression of the dijet cross section at the energies beyond the crossover energy $b^{-1}$, which may be the only detectable signature of quantum black hole formation.

An alternative model for space-time at short distances is that of space-time foam. In that case we can assume that the quantum black hole forms as previously described, but decays in a pair of jets emitted back-to-back in a random direction, according to the stochastic fluctuations of space-time at short distances. In this model we expect modification of dijet angular distributions above the formation threshold of QBHs, which would resemble that for a compositeness signal (i.e., an excess of centrally produced jets, unlike mostly forward jets from the $t$-channel exchange dominating pure QCD dijet production mechanism).

As in  \cite{Calmet:2008dg}, we assume that processes involving QBHs conserve QCD and $U(1)$ charges since local gauge symmetries are not violated by gravity. Note that we make no similar assumption about global charges. For example, B-L, Lorentz invariance, or flavor number do not have to be preserved in quantum gravitational interaction. Note that the Lorentz invariance is expected to be violated if the dimensionality of space-time shrinks at short distances. 
The total cross section for black hole production in our model is quite large. We find $\sigma(p+p \to  \ \mbox{QBH} ) \approx 140$ pb for a 14 TeV LHC. For the start-up LHC operating energy of 7 TeV, the cross section is still sizable: $\approx 18$~pb. The different possible final states allowed by gauge invariance have been discussed in \cite{Calmet:2008dg}.  Quantum black holes produced at a proton-proton collider can be classified according to representations of $SU(3)_{\mbox{color}}$ and $U(1)_{\mbox{QED}}$. This enables us to make predictions for production of certain exotic final states, shown in Table~1. Note that we are allowing transitions, which violate Lorentz symmetry, such as, e.g., those described by effective operators of the type $q q \bar q  \gamma$. Since, in our scenario, the dimensionality of space-time shrinks at high energy, Lorentz invariance is not expected to hold above the dimensional crossover. In deriving the branching fractions listed in Table~1, we assume that all the transitions not explicitly protected by a gauge symmetry take place. For example, we include transitions described by effective operators of the type $qqql$ or $q\bar q e^+ \mu^-$. 

The branching fractions in Table~1  are model dependent. We assume that there are about hundred possible degrees of freedom accessible to QBHs in models with large extra dimensions. In the case of a large hidden sector, there are $\sim 10^{33}$ degrees of freedom. Neutral QBHs will thus decay predominantly into the hidden sector. Hence the branching fraction of QBHs in that scenario to two oppositely charged leptons is essentially zero. On the other hand, charged QBHs in this scenario, will still predominantly decay into about hundred possible degrees of freedom of the standard model, because of gauge invariance. Independently of the model, most QBHs produced at the LHC carry QCD and QED charges since there are produced in collisions of quarks and gluons. These QBHs will decay mainly to colored particles of the standard model and will thus lead to dijets events. Table~1 focuses on smocking-gun exotic signatures for QBHs, with potentially little background from standard model physics.


\begin{table}[tb]\label{tableQBH}
\begin{center}
\begin{tabular}{|c|c|}
\hline
 &Branching fractions for interesting decay modes \\
\hline
$\mbox{Br}(p+p \to \mbox{QBH}^{4/3}_{\bar{3}} \to l^+ + \bar d )$ &  $\sim 0.14 \%$ \\
\hline
$\mbox{Br}(p+p \to \mbox{QBH}^{-2/3}_{\bar{3}} \to l^- + \bar d )$& $\sim 0.04\%$ \\
\hline
$\mbox{Br}(p+p \to \mbox{QBH}^{1/3}_{\bar{3}} \to \nu_i + \bar d )$& $\sim 0.1 \%$ \\ \hline
$\mbox{Br}(p+p \to \mbox{QBH}^{-2/3}_{\bar{3}} \to \nu_i + \bar u )$& $\sim 0.04 \%$ \\ \hline
$\mbox{Br}(p+p \to \mbox{QBH}^{-2/3}_{\bar{3}} \to \gamma+ \bar u )$& $\sim 0.04 \%$ \\\hline
$\mbox{Br}(p+p \to \mbox{QBH}^{1/3}_{\bar{3}} \to \gamma + \bar d )$& $\sim 0.1 \%$ \\ \hline
$\mbox{Br}(p+p  \to \mbox{QBH}^0_1$ $\to e^+ + \mu^-)$&$\sim$ 0 to 0.006 \% \\\hline
\end{tabular}
\end{center}
\caption{Some possible exotic final states in quantum black hole decays. The branching fraction $\mbox{Br}(p+p \to \mbox{QBH} \to Z+\mbox{jet}) = 3/2 \times \mbox{Br}(p+p \to \mbox{QBH} \to \gamma+\mbox{jet})$. Note that the final state for neutral black holes depends on the model for strong gravity under consideration. In a large hidden sector model, neutral black holes decay predominantly into the hidden sector. In models with large extra dimensions, these black holes can have sizable decay into two charged leptons., e.g., $e^+ \mu^-$.}
\end{table}

Another potential model is that without large extra dimensions or a large hidden sector, but with space-time becoming two-dimensional at short distances as discussed in \cite{Anchordoqui:2010er}. In this model semi-classical black holes will not form in a collision of two partons at TeV energies; however, if the parameters of the action (\ref{action2d}) are such that 1+1 gravity is strong, quantum black holes could still form. In that case, the impact parameter is of the order $b\sim (1\ \mbox{TeV})^{-1}$ and the cross section could depend on the parameters of the fundamental action (\ref{action2d}). We thus would expect $\sigma_4=\sigma_2 b^{-2} = f(\sqrt{\hat s}/\Lambda)) b^{-2}$. The function $f$ is unknown and it is thus difficult to make numerical predictions. However, if as assumed in \cite{Anchordoqui:2010er}, lowering the dimensionality of space-time solves the hierarchy problem, the cross sections for quantum black holes would be still of the order of $b^2 \sim 1\ \mbox{TeV}^{-2}$ and thus comparable in magnitude to those obtained above.

Note that in all of the above scenarios, which allow formation of semi-classical and classical black holes, their physics is not expected to  be affected since their Schwarzschild radii are much larger than the typical scale $b$ where the dimensionality of space-time changes.

{\it Conclusions:} We have considered quantum black holes at the LHC in the context of models of quantum gravity, which are lower-dimensional at short distances. Different scenarios are discussed. The first ones are within the context of well-studied models with a large extra-dimensional volume or a large hidden sector. In these scenarios, it is possible to either suppress quantum black hole production completely, or on the contrary produce them copiously at the LHC. We find that the cross sections for small non-thermal black holes can be quite different from those considered previously in the literature. The cross section can be quite large and does not depend on the black hole mass. The angular distribution of the final state particles would give important insight into the nature of space-time at short distance and will help to differentiate different models of space-time foam. The final state in the black hole decay will allow to determine the model at the origin of strong gravitational physics in the TeV range and help to probe the symmetries of quantum gravity. Clearly, the search strategies for mini black holes at the LHC have to be adapted to be able to detect the QBHs described in the paper. Finally, if space-time becomes 1+1-dimensional, as recently proposed to address the hierarchy problem of the standard model, gravity could become strongly coupled when space-time becomes 1+1-dimensional. In that case, only quantum black holes would be produced in particle collisions.

{\it Acknowledgments:}

The work of XC is supported in part by the European Cooperation in Science and Technology (COST) action MP0905 ``Black Holes in a Violent Universe". The work of GL is partially supported by the US Department of Energy, under Grant No. DE-FG02-91ER40688. We thank the organizers of the Experimental Searches for Quantum Gravity Workshop ESQG 2010, where this work has originated.



\bigskip


\end{document}